\documentclass[aoas,nameyear,seceqn,rotating,dvips]{arximspdf}
\usepackage{dcolumn}
\usepackage{graphicx}

\doi{10.1214/10-AOAS393}
\volume{5}
\issue{2A}
\pubyear{2011}
\firstpage{994}
\lastpage{1019}

\makeatletter
\newcolumntype{d}[1]{D{.}{.}{#1}}
\newproclaim{rem}{Remarks}
\newtheorem{theo}{Theorem}
\newtheorem{corollary}{Corollary}
\makeatother

\begin{document}
\begin{frontmatter}

\title{An adaptively weighted statistic for detecting differential gene expression when combining multiple transcriptomic studies}
\runtitle{Adaptively weighted statistic}

\begin{aug}
\author[A]{\fnms{Jia} \snm{Li}\ead[label=e2]{jli3@hfhs.org}}
\and
\author[A]{\fnms{George C.} \snm{Tseng}\corref{}\thanksref{t1}\ead[label=e1]{ctseng@pitt.edu}}

\thankstext{t1}{Supported in part by NIH (KL2 RR024154-02) and the University
of Pittsburgh (Central Research Development Fund, CRDF; Competitive
Medical Research Fund, CMRF).}
\runauthor{J. Li and G. C. Tseng}
\affiliation{University of Pittsburgh}
\address[A]{Department of Biostatistics\\
University of Pittsburgh\\
Pittsburgh, Pensylvania\\
USA\\
\printead{e2}\\
\phantom{E-mail:\ }\printead*{e1}} %adresu isvedimo komanda gale!
\end{aug}

% HISTORY:
\received{\smonth{6} \syear{2009}}
\revised{\smonth{7} \syear{2010}}

% ABSTRACT
\begin{abstract}
Global expression analyses using microarray technologies are becoming
more common in genomic research, therefore, new statistical challenges
associated with combining information from multiple studies must be
addressed. In this paper we will describe our proposal for an
adaptively weighted (AW) statistic to combine multiple genomic studies
for detecting differentially expressed genes. We will also present our
results from comparisons of our proposed AW statistic to Fisher's
equally weighted (EW), Tippett's minimum $p$-value (minP) and Pearson's
(PR) statistics. Due to the absence of a uniformly powerful test, we
used a simplified Gaussian scenario to compare the four methods. Our AW
statistic consistently produced the best or near-best power for a range
of alternative hypotheses. AW-obtained weights also have the additional
advantage of filtering discordant biomarkers and providing natural
detected gene categories for further biological investigation. Here we
will demonstrate the superior performance of our proposed AW statistic
based on a mix of power analyses, simulations and applications using
data sets for multi-tissue energy metabolism mouse, multi-lab prostate
cancer and lung cancer.
\end{abstract}

% KEYWORDS

\end{frontmatter}

%s1 ###
\section{Introduction}\label{sec:intro}
Integrating results from multiple biological studies is now considered
commonplace, with significance levels and effect sizes often used in
meta-analyses. Random effects models which models effect sizes are
frequently used to address variation in sampling schemes. Differences
in data structures and statistical hypotheses are common in multiple
applications, making direct combinations of effect sizes difficult or
impossible. It is more feasible to combine the transformed probability
integrals of test statistics (usually $p$-values), since the procedure
is only dependent on the significance values of individual tests
instead of on underlying data structures. Fisher's (\citeyear{Fisherh1932})
well-known method of this type involves the log-transformation of
$p$-values to Chi-square\vspace*{1pt} scores and the equally-weighted summation:
$V^{\mathrm{EW}}=-\sum_{k=1}^K \log(p_k)$, where $K$ studies are combined and
$p_k$ is the $p$-value of study $k$, $1\leq k\leq K$. Assuming
independence among studies and $p$-values calculated from correct null
distributions in each study, $2V^{\mathrm{EW}}$ follows a Chi-square
distribution with $2K$ degrees of freedom under the null hypothesis.
Previously considered other transformations include inverse normal
[Stouffer et al. (\citeyear{Stoufferh1949})], logit [Lancaster
(\citeyear{Lancasterh1961})] and inverse Chi-square transformation with varying degrees of
freedom [George (\citeyear{Georgeh1977})], among many others. Although Fisher's method is
not the most uniformly powerful, it does exhibit good power for a wide
range of conditions. It is also recognized for its asymptotically
Bahadur optimal (ABO) characteristic, with multiple studies having the
same effect size for alternative hypotheses [Littell and Folks (\citeyear{Littleh1971,Littleh1973})]. Different weights or variations of Fisher's statistic have also
been considered. Good (\citeyear{Goodh1955}) suggested using unequal weights for
individual studies in which weights are determined by decisions made by
subject experts. More recently, Olkin and Saner (\citeyear{OlkinandSanerh2001}) have proposed a
trimmed version of Fisher's statistic to remove the potential effects
of aberrant extremes. Another well-known method in the category of
combining $p$-values is Tippett's (\citeyear{Tippetth1931}) minimum $p$-value statistic
(minP): $V^{\mathrm{minP}}=\min_{1\leq k \leq K}p_k$. Wilkinson (\citeyear{Wilkinsonh1951})
generalized Tippett's procedure to a more robust $r$th smallest $p$-value,
in which $V^{\mathrm{maxP}}=\max_{1\leq k \leq K}p_k$ (maxP) is widely used.
Note that minP and maxP statistics align with Roy's (\citeyear{Royh1953})
union--intersection test and Berger's (\citeyear{Bergerh1982}) intersection--union test,
respectively. For comprehensive reviews and comparisons of various
meta-analysis approaches, see Hedges and Olkin (\citeyear{HedgesandOlkinh1985}) and Cousins (\citeyear{Cousinsh2007}).

Microarray supports the examination of the expression of thousands of
genes in parallel. As microarray experiments become more mature and
common, it has become increasingly important to integrate homogeneous
experimental data sets from multiple laboratories and experimental
techniques. In contrast to traditional epidemiological or
evidence-based medical studies, the process of monitoring the
expression for thousands of genes simultaneously presents many
challenges to integrative analysis. In the current biological
literature, the term meta-analysis refers to the widespread use of
naive intersection/union operations or vote counting on lists of
differentially expressed genes obtained from individual studies using
certain criteria---for instance, False Discovery Rate${}\leq 0.05$
[Borovecki et al. (\citeyear{Boroveckih2005}); Cardoso et al.
(\citeyear{Cardosoh2007}); Pirooznia, Nagarajan and Deng
(\citeyear{Piroozniah2007}); Segal et al. (\citeyear{Segalh2004}),
among many others]. Intersections are too conservative and unions
insufficiently conservative, especially as the value of $K$ increases.

More sophisticated meta-analysis methods can be divided into two
traditions, the first being the use of a summary statistic---that is, a
combination of statistics from individual studies for each gene being
considered, adjusted for multiple comparisons. In many situations, this
type of method is an extension of traditional meta-analysis methods.
For example, Rhodes et al. (\citeyear{Rhodesh2002}), who were the
first to apply Fisher's method to microarray data, later introduced a
weighted average of test statistics from individual tests, with weights
determined by study sample sizes [Ghosh et al. (\citeyear{Ghoshh2003})]. Moreau et al. (\citeyear{Moreauh2003}) made use of Tippett's
minimum $p$-value. A more robust statistic is Wilkinson's $r$th smallest
$p$-value, in which maximum $p$-value can be applied to the meta-analysis
of microarray studies. Owen (\citeyear{Owenh2009}) reintroduced Pearson's (\citeyear{Pearsonh1934}) method
and applied it to the AGEMAP project. He defined a test statistic as
the maximum of Fisher's combination of left-sided and right-sided
$p$-values. All of these methods combine statistical significance. Note
that when no gene effect exists, the $p$-value is uniformly distributed.
Accordingly, combining the significance of independent tests is
sometimes called omnibus or nonparametric. When studies have similar
design and measure the outcomes in similar ways, combining effect sizes
is usually preferred to combining significance. Choi et
al. (\citeyear{Choih2003}) used weighted estimate for individual genes based on the
random effects model (REM) under Gaussian assumptions, and discussed
the details of a Bayesian formulation for the REM model. Hu, Greenwood
and Beyene (\citeyear{Huh2005}) developed a quality measure for each gene in
individual studies, incorporating a quality index as a weight in the
REM model. Hong et al. (\citeyear{Hongh2006}) proposed a robust
rank-based approach for meta-analysis. Choi et al.
(\citeyear{Choih2007}) introduced a latent variable approach.

The second meta-analysis tradition is Bayesian---for example, Choi et
al.'s (\citeyear{Choih2003}) Bayesian version for REM, which models the effect sizes.
Similar Bayesian hierarchical models have been suggested by Tseng et al. (\citeyear{Tsengh2001}) and Conlon, Song and Liu (\citeyear{Conlonh2006}) for
incorporating different levels of replicates information in cDNA
microarray experiments. Conlon, Song and Liu (\citeyear{Conlonh2007}) refer to these
models as Bayesian probability integration (PI) models, and have
introduced a Bayesian standardized expression integration (SEI) model.
Instead of modeling study specific means separately (PI model), SEI
models them as samples from a normal distribution, thus producing
overall mean and inter-study variation. Shen, Ghosh and Chinnaiyan
(\citeyear{Shenh2004}) and Choi et al. (\citeyear{Choih2007}) used a Bayesian mixture model to rescale
the individual data set and then combined all data sets for an ordinary
gene expression analysis.

The structure for the rest of this paper is as follows: in Section
\ref{sec:HS1_HS2} we describe two complementary hypothesis settings for
detecting study-invariant and study-specific biomarkers: $\mathit{HS}_A$ and
$\mathit{HS}_B$. In Section~\ref{sec:AW} we present our proposal for an
adaptively weighted (AW) statistic for meta-analyses of genomic
studies, including detailed descriptions of the AW statistic algorithm
and a permutation test for combining multiple studies. In Section
\ref{sec:application} we discuss a simulation test of our proposed
method, using data sets from studies of a multi-tissue energy
metabolism mouse model, prostate cancer and lung cancer; we then
compare our results with those produced by three other commonly used
methods. In Section~\ref{sec:compare} we demonstrate the admissibility
and power of our proposed AW test under a Gaussian assumption, and in
Section~\ref{sec:discussion} we summarize its statistical advantages
and limitations.

%s2 ###
\section{Two major complementary hypothesis settings}\label{sec:HS1_HS2}

To our knowledge, no comprehensive evaluations for the above-described
meta-analysis methods have been performed, primarily due to a lack of
rigorous formulation of statistical hypotheses. Here we will consider a
meta-analysis of $D_1, D_2,\ldots,D_K$ gene expression profiles
studies. $x_{kgs}$ is the gene expression intensity of gene $g$ and
sample $s$ in study $k$, with samples $s=1,\ldots,n_k$ belonging to a
control group (e.g., normal samples) and $s=n_k+1,\ldots,n_k+m_k$
belonging to the diseased group (e.g., cancer samples). Normally a null
hypothesis for each gene $g$ is considered as
\[
H_0\dvtx
\theta_{g1}=\cdots=\theta_{gK}=0,
\]
where $\theta_{gk}$ represents the
gene effect of gene $g$ and study $k$. Building on Birnbaum's (\citeyear{Birnbaumh1954})
work, the complementary hypothesis settings ($\mathit{HS}_A$ and $\mathit{HS}_B$) are
dependent upon the nature of the experiment in which the gene effects
($\theta_{gk}$) are obtained:
\begin{eqnarray*}
&&
\mathit{HS}_A\dvtx \{H_0 \mbox{ versus } H_A\dvtx \theta_{gk}\neq0, \forall 1\leq k\leq K\},
\\
&&
\mathit{HS}_B\dvtx \{H_0 \mbox{ versus } H_B\dvtx \mbox{at least one } \theta_{gk}\neq 0, 1\leq k \leq K\}.
\end{eqnarray*}
It is possible to use different methods to explicitly or implicitly
consider different subsets or variations of the two alternative
hypotheses:
\begin{eqnarray*}
&&
\mathit{HS}_{A1}\dvtx \{H_0\mbox{ versus }
H_{A1}\dvtx\theta_g=\theta_{g1}=\cdots=\theta_{gK}\neq0\},
\\
&&
\mathit{HS}_{A2}\dvtx\{ H_0 \mbox{ versus } H_{A2}\dvtx \theta_g\neq 0,\theta_{gk}\sim N(\theta_g,
\tau^2)\},
\\
&&
\mathit{HS}_{Bh}\dvtx\{H_0 \mbox{ versus } H_{Bh}\dvtx \sum_{k=1}^K I(\theta_{gk}\neq 0)=h\mbox{ }(1\leq h\leq
K)\}
\\
&&\qquad
[I(\cdot) \mbox{ is an indicator function that}
\\
&&\qquad\hphantom{[}\mbox{equals 1 when statement true and 0 otherwise]},
\\
&&
\mathit{HS}_{Bh'}\dvtx\Biggl\{H_0 \mbox{ versus }  H_{Bh'}\dvtx \sum_{k=1}^K I(\theta_{gk}\neq 0)=h
\\
&&\hphantom{\mathit{HS}_{Bh'}\dvtx\Biggl\{}\mbox{ and } \theta_{gk}=\theta_g \mbox{ if }\theta_{gk}\neq 0\ (1\leq h\leq
K)\Biggr\}.
\end{eqnarray*}

Without danger of confusion, here we will use $H_A$ notation to denote
the parameter space of the corresponding alternative hypothesis. It is
clearly seen that $H_A\subset H_B$. However, they represent two
families of complementary interpretations in applications. Under $H_A$,
gene $g$ is identified only when it is differentially expressed in all
studies. Under $H_B$, gene $g$ is selected only if it is differentially
expressed in one or more studies. Note that $H_{A1}\subset H_A$,
representing an equal fixed effect model. $H_{A2}$ represents a random
effects model for a similar $H_A$ purpose, while $H_{A2}\not\subseteq
H_A$ in general. Note also that $H_B=\bigcup_{1\leq h\leq K} H_{Bh}$,
$H_{Bh'}\subset H_{Bh}$ ($1\leq h\leq K$) and $H_{BK'}=H_{A1}$.

%t1 ###
\begin{sidewaystable}
\tablewidth=551pt
\caption{Three sets of microarray studies for meta-analyses. (BF---brown fat; Liv---liver; Ht---heart; WT---wild type (VLCAD $+$/$+$); VLCAD---VLCAD $-$/$-$; N---normal; T---tumor; AC---adenocarcinomas)}
\label{three-datasets}
\begin{tabular*}{\textwidth}{@{\extracolsep{\fill}}lld{1.2}d{1.2}d{1.2}ld{2.2}d{2.2}d{2.2}ld{3.2}d{2.2}d{2.2}@{}}
\hline
\textbf{(A)}&\multicolumn{4}{c}{\textbf{Mouse energy metabolism}}&\multicolumn{4}{c}{\textbf{Prostate cancer studies}}&\multicolumn{4}{c@{}}{\textbf{Lung cancer studies}}\\[-5pt]
&\multicolumn{4}{c}{\hrulefill}&\multicolumn{4}{c}{\hrulefill}&\multicolumn{4}{c@{}}{\hrulefill}\\
&&\multicolumn{1}{c}{\textbf{BF}}&\multicolumn{1}{c}{\textbf{Liv}}&\multicolumn{1}{c}{\textbf{Ht}}&&\multicolumn{1}{c}{\textbf{Dhan}}&\multicolumn{1}{c}{\textbf{Luo}}&\multicolumn{1}{c}{\textbf{Wels}}&&\multicolumn{1}{c}{\textbf{Bhat}}&\multicolumn{1}{c}{\textbf{Beer}}&\multicolumn{1}{c@{}}{\textbf{Garb}}\\[-5pt]
&\multicolumn{4}{c}{\hrulefill}&\multicolumn{4}{c}{\hrulefill}&\multicolumn{4}{c@{}}{\hrulefill}\\
&WT & 4 & 4 & 3 & N & 19 & 9 & 9 & N & 17 & 10 & 5  \\
& VLCAD & 4 & 4 & 4 & T & 14 & 16 & 25 & AC & 134 & 86 & 39\\[6pt]
(B)&&&&&&&&&&&&\\
$\mathit{HS}_A$&\multicolumn{4}{c}{Of biological interest}&\multicolumn{4}{c}{Of biological interest}&\multicolumn{4}{c@{}}{Of biological interest}\\
$\mathit{HS}_B$&\multicolumn{4}{c}{Of biological interest}&\multicolumn{4}{c}{Of less biological interest}&\multicolumn{4}{c@{}}{Of less biological interest}\\
&&&&&\multicolumn{4}{c}{but of more technical interest}&\multicolumn{4}{c}{but of more technical interest}\\[6pt]
(C)&&&&&&&&&&&&\\
&BF & 1 & 0.06 &0.04  & Dhan & 1 &0.05  & 0.09  & Bhat &1  &0.33  &0.22   \\
&Liv &0.06  &1  &0.03 & Luo & 0.05  & 1 & 0.09  & Beer &0.33  & 1 &0.15   \\
&Ht &0.04  &0.03 & 1 & Wels & 0.09  & 0.09  & 1 & Garb &0.22  &0.15  & 1  \\
\hline
\end{tabular*}
\end{sidewaystable}

From a biological standpoint, experimental design and meta-analysis
objectives determine biomarker lists of interest. To illustrate this
idea, we will use three sets of microarray studies for meta-analyses.
The first set consists of two mouse genotypes, wild type (VLCAD $+$/$+$)
and VLCAD deficient (VLCAD $-$/$-$), with four mice in each genotype group
(VLCAD is associated with a childhood metabolism disorder). Brown fat,
liver and heart tissue samples were collected from each of the eight
individual mice and used for microarray experiments designed to study
global expression changes in the knock-out of VLCAD (Table
\ref{three-datasets}, left). Given the experimental design, a biomarker
list of interest might consist of those genes that are consistently
expressed in all tissue samples from both wild type and VLCAD-deficient
mice. This type of tissue-invariant (or study-invariant) biomarker list
can be loosely defined as $G_A$, with analysis based on the alternative
hypothesis family of $H_A$. However, it is reasonable to assume that
tissue-specific physiology triggers tissue-dependent responses, with
pools of differentially expressed genes being confounded to the tissues
in question. Such a hypothesis would focus on signature genes that are
differentially expressed in subsets of one or more tissues---an
analysis that corresponds to the $H_B$ alternative hypothesis family.
Hereafter we will use the term $G_B$ when addressing such
tissue-specific or study-specific biomarker lists. In the second study
set, microarray comparisons of normal versus prostate tumor tissues
were performed by three different research teams: Dhanasekaran et al. (\citeyear{Dhanasekaranh2001}), Luo et al. (\citeyear{Luoh2001})
and Welsh et al. (\citeyear{Welshh2001}) (Table~\ref{three-datasets}).
The $G_A$ study-invariant biomarker list is clearly of greater
biological interest in this situation, since many of the $G_B$
study-specific biomarkers represent experimental and technical
discrepancies between studies, possibly due to sample population
heterogeneity, gene matching errors or differences in experimental
protocols. Further investigation of study-specific biomarkers may
provide technical insights to experimental design features without
providing biological insights to the disease of interest. The third set
of microarray studies [Bhattacharjee et al.
(\citeyear{Bhath2001}); Beer et al. (\citeyear{Beerh2002}); Garber et al. (\citeyear{Garberh2001})] included analyses of lung cancer samples and a
comparison of normal versus adenocarcinoma samples. Table
\ref{three-datasets}C shows the pair-wise integrative correlation
coefficients [Parmigiani et al. (\citeyear{Parmigianih2004})]
in each of the three examples. A~review of past meta-analyses reveals
that lung cancer studies generally have larger samples, greater
homogeneity and better data quality than prostate cancer studies,
especially in terms of biomarker detection and classification analysis.

%t2 ###
\begin{table}
\tablewidth=318pt
\caption{Meta-analysis methods, corresponding hypothesis settings and targeted types of biomarker list}
\label{methods_table}
\begin{tabular}{@{}l@{\hspace{3pt}}ccc@{}}
\hline
 &  & \textbf{Alternative} & \textbf{Targeted} \\
\textbf{Methods}&\textbf{Abbreviation}& \textbf{hypothesis} & \textbf{biomarker list}\\
\hline
Fisher [equally weighted sum of  & EW & $H_B$ & $G_B$ \\
log($p$-values)]&&&\\
Tippett (minimum $p$-value) & minP & $H_B$ & $G_B$ \\
Pearson (maximum of Fisher's left- & PR & $H_B$ & $G_B$\\
sided and right-sided score)&&&\\
Li and Tseng [adaptively weighted  & AW & $H_B$ & $G_B$\\
sum of log($p$-values)]&&&\\[3pt]
Wilkinson (maximum $p$-value) & maxP & $H_A$ & $G_A$\\
Choi (\citeyear{Choih2003}); Shen (\citeyear{Shenh2004}); Choi (\citeyear{Choih2007})  & REM & $H_{A2}$ & $G_A$\\
(random effects model)&&&\\
Conlon (\citeyear{Conlonh2006}) (PI Bayesian approach) & PI & NA & $G_A$\\
Conlon (\citeyear{Conlonh2007}) (SEI Bayesian approach) & SEI & NA & $G_A$\\
\hline
\end{tabular}
\end{table}

Table~\ref{methods_table} presents a list of commonly used
meta-analysis methods for microarray studies, their corresponding
alternative hypotheses and targeted biomarkers. While both Bayesian SEI
and PI methods tend to detect $G_A$-type biomarkers across studies, the
Bayesian concept does not involve hypothesis testing. Note that
different approaches have distinctly different advantages and
disadvantages in terms of parameter space subsets in alternative
hypotheses, even though two methods may be designed for the same
hypothesis. For example, to detect $G_A$ genes, PI performs better than
SEI for genes that have a high mean effect in one study but low mean
effect in another. According to Laughin (\citeyear{Loughinh2004}), maxP is generally
under-powered, but performs well when all $\theta_{gk}$ values are
nonzero and roughly the same. As we will show in Section
\ref{sec:compare}, EW, minP, PR and AW are all admissible for detecting
$G_B$ genes. For $H_{Bh}$, EW tends to be more powerful when $h$ is
large and closer to $K$. Little and Folks proved that EW is
asymptotically ABO when detecting $G_A$-type genes under under
$H_{BK'}$ (i.e., $H_{A1}$), even though the EW statistic is targeted
toward general $H_B$. In contrast, minP is more powerful in detecting
genes under $H_{Bh}$ when $h$ is small.

From this point forward, our focus will be on the $H_B$ alternative
hypothesis. In the following section we will describe our proposal for
an adaptively weighted statistic (AW), and, in Section
\ref{sec:compare}, we will demonstrate its robustness and near-optimal
power for alternative hypotheses at either extreme (i.e., when $h$ is
close to $K$ or close to 1 in $H_{Bh'}$). We will also give examples of
situations in which AW outperforms EW and minP in intermediate
scenarios. AW is capable of distinguishing $G_A$ and $G_B\backslash
G_A$ genes in a manner that indicates in which study or studies
individual biomarkers are differentially expressed---information
considered useful for post-meta-analysis investigations.

%s3 ###
\section{Adaptively-weighted statistic}\label{sec:AW}
When integrating multiple genomic studies, expression of some important
biomarkers may be altered in a study-specific manner (consider $H_B$).
To uncover altered gene expression patterns across studies, we start
with the following weighted statistic:
%e3.1 ###
\begin{equation}
U_g(w_g)=-\sum_{k=1}^Kw_{gk} \log(p_{gk}),
\end{equation}
where $p_{gk}$ is the $p$-value of gene $g$ in study $k$, $w_k$ is the
weight assigned to the $k$th study and $w_g=(w_{g1},\ldots,w_{gK})$.
Under the null hypothesis that $\theta_{gk}=0$ $\forall k$, the $p$-value
of the observed weighted statistic, $p_U(u_g(w_g))$, can be obtained
for a given gene $g$ and weight $w_g$ (see below for detailed
permutation algorithm to calculate the $p$-value). The
adaptively-weighted statistic is defined as the minimal $p$-value among
all possible weights:
%e3.2 ###
\begin{equation}
V^{\mathrm{AW}}_g=\min_{w_g \in W}p_U(u_g(w_g)),
\end{equation}
where $u_g(w)$ is the observed statistic for $U_g(w)$, and $W$ is a
prespecified search space. Our choice of search space in this paper is
$W=\{w\mid w_i\in \{0,1\}\}$, which results in an affordable computation of
$O(2^K-1)$ based on the norm of $K\leq 10$ in a microarray
meta-analysis.

The resulting weight reflects a natural biological interpretation of
whether or not a study contributes to the statistical significance of a
gene. Note that the AW statistic is inadequate for traditional
meta-analysis in epidemiological or evidence-based medicine research.
The AW selection procedure will introduce selection bias toward studies
with concordant significant effects. However, integrative analysis of
genomic studies represents a different situation: usually the primary
goal is to screen and identify the most probable gene markers, given
data meant to facilitate future investigation. As we will show in
Section~\ref{sec:application}, the weight vector, $w_g^*=\arg\min_{w_g
\in W}p_U(u_g(w_g))$, actually serves as a convenient basis for gene
categorization in follow-up biological interpretations and
explorations.

Below we illustrate the detailed procedure for AW when applied to
combined genomic studies. If assuming $p_{gk}\sim \operatorname{Unif}[0,1]$ under the
null hypothesis, $U_g(w_g)\sim \operatorname{Gamma}(\sum_{k=1}^K w_{gk}, 1)$ and
inference of the AW statistic can be performed on this basis. Such a
uniform $p$-value assumption is, however, usually not true in real
applications. Alternatively, a permutation test is performed below to
assess the statistical significance and the false discovery rate (FDR)
is controlled at 5\%. For the applications in Section
\ref{sec:application}, the EW, minP, maxP and PR methods are performed
using a similar permutation test.

\begin{enumerate}[III.]
\item[I.]  Study-wise $p$-value calculation before meta-analysis:
\begin{enumerate}
\item[(1)] Compute the penalized $t$-statistics, $t_{gk}$, for gene $g$
and study $k$ [Efron et al. (\citeyear{Efronh2001}); Tusher, Tibshirani
and Chu (\citeyear{Tusherh2001})].
\item[(2)] Permute group labels in
each study for $B$ times, and similarly calculate the permuted
statistics, $t_{gk}^{(b)}$, where $1\leq g \leq G, 1\leq k \leq K,1\leq
b \leq B$.
\item[(3)] Estimate the $p$-value of $t_{gk}$ as
$p_{gk}=(\sum_{b=1}^B \sum_{g'=1}^G I(t_{g'k}^{(b)}\in
R(t_{gk})))/\break
(B\cdot G)$, where $R(t_{gk})$ is the rejection region given the
threshold $t_{gk}$. Similarly, given $t_{gk}^{(b)}$, compute
$p_{gk}^{(b)}=(\sum_{b'=1}^B \sum_{g'=1}^G I(t_{g'k}^{(b')}\in
R(t_{gk}^{(b)})))/\break(B\cdot G)$.\vspace*{1pt}
\end{enumerate}
\item[II.] Calculate AW statistic:
\begin{enumerate}[(3)]
\item[(1)] Given a weight $w_g=(w_{g1},\ldots,w_{gK})$, the weighted statistic is defined as $u_g(w_g)=-\sum_{k=1}^Kw_{gk} \log(p_{gk})$ for gene $g$.
Define $u_g^{(b)}(w_g)=-\sum_{k=1}^Kw_{gk} \log(p_{gk}^{(b)})$.
\item[(2)] Estimate the $p$-value of the observed $u_g(w_g)$ as
\[
\displaystyle p_U(u_g(w_g))=\frac{\sum_{b=1}^B \sum_{g'=1}^G I\{u_{g'}^{(b)}(w_g)
\geq u_g(w_g)\}}{B\cdot G}.
\]
Similarly compute
\[
\displaystyle p_U\bigl(u_g^{(b)}(w_g)\bigr)=\frac{ \sum_{b'=1}^B \sum_{g'=1}^G I\{u_{g'}^{(b')}(w_g) \geq u_g^{(b)}(w_g)\}}{B\cdot
G}.
\]
\item[(3)] Based on II(1) and II(2), calculate the optimal weight as
\[
w_g^*=\arg\min_{w_g\in W}p_U(u_g(w_g))
\]
and, similarly,
\[
w_g^{(b)*}=\arg\min_{w_g\in W} p_U\bigl(u_g^{(b)}(w_g)\bigr).
\]
Define the AW statistic $V_g$ as the $p$-value of the adaptively weighted statistic: $V_g=p_U(u_g(w_g^*))$. Similarly, $V_g^{(b)}=p_U(u_g^{(b)}(w_g^{(b)*}))$.
\end{enumerate}
\item[III.] Assess $p$-values and $q$-values of the AW statistic---$V_g$:
\begin{enumerate}[(3)]
\item[(1)] The $p$-value of $V_g$ is calculated as
\[
\displaystyle p_V(V_g)=\frac{ \sum_{b=1}^B \sum_{g'=1}^G I\{V_{g'}^{(b)}\leq V_g\}}{B\cdot G}.
\]
\item[(2)] Estimate $\pi_0$, the proportion of null genes, as
\[
\displaystyle \widehat{\pi}_0=\frac{\sum_{g=1}^G I\{p_V(V_g)\in A\}}{G\cdot \ell(A)}
\]
[Storey (\citeyear{s2002})].
Normally we choose $A=[0.5,1]$ and $\ell(A)=0.5$.
\item[(3)] Estimate the $q$-value for each gene as
\[
\displaystyle q(V_g)=\frac{{\widehat{\pi}_0} \sum_{b=1}^B \sum_{g'=1}^G I\{V_{g'}^{(b)}\leq V_g\}} {B\sum_{g'=1}^G I\{V_{g'}\leq V_g\}}.
\]
The detected gene list is $G^{\mathrm{AW}}=\{g\dvtx q_V(V_g)\leq 0.05\}$.
\end{enumerate}
\item[IV.] Distinguish concordant and discordant genes (recommended):
Split the detected gene list $G^{\mathrm{AW}}$ into concordant and discordant
gene lists. By controlling the false discovery rate (FDR) at 5\%,
detected genes with concordant regulation direction across contributing
studies are denoted as $G^{\mathrm{AW}}_{\mathrm{concordant}}=\{g\dvtx q(V_g)\leq
0.05$ and $|\sum_{k=1}^K \operatorname{sgn}(t_{gk})\cdot
w^*_{gk}|=\sum_{k=1}^K w^*_{gk}\}$, where $\operatorname{sgn}(\cdot)$ is the sign
function that takes value 1 when positive and $-$1 when negative. The
discordant gene list is $G^{\mathrm{AW}}_{\mathrm{discordant}}=G^{\mathrm{AW}}\backslash
G^{\mathrm{AW}}_{\mathrm{concordant}}$.
\end{enumerate}

\begin{rem*}
\begin{enumerate}
\item[1.] For the application of EW and the minP, maxP and PR method,
steps
II(1)--II(3) can be skipped. Alternatively, the test statistics are
modified as $V_g=-\sum_{k=1}^K \log(p_{gk})$ for EW; $V_g=\min_{1\leq k\leq K} p_{gk}$ for minP; $V_g=\max_{1\leq k\leq
K} p_{gk}$ for maxP and $V_g=\max(-\sum_{k=1}^K
\log(\tilde{p}_{gk}), -\sum_{k=1}^K\log(1-\tilde{p}_{gk}))$ for PR,
where $\tilde{p}_{gk}$ is the one-sided $p$-value for gene $g$ in study
$k$.
\item[2.] The I--III sequence provides an algorithm for a general
framework. Both statistics $t_{gk}$ and rejection region $R(t_{gk})$
can be replaced, depending on the experimental design and hypothesis.
For example, the $F$-statistic can be used when multiple groups of
samples are available in each study under consideration.
\item[3.] When conducting comparisons of two groups and applying the
moderated $t$-statistic, genes detected under the general framework (the
I--III sequence) may contain discordant genes---for instance, a gene
up-regulated in one study and down-regulated in another; the addition
of step IV provides further filtering. In some applications, a
researcher may want to scrutinize the discordant gene list to verify
whether the discordance reflects actual biological discrepancy across
studies (e.g., different tissues or patient populations) or artificial
errors (e.g., mistakes in gene annotation). For EW and minP there is no
direct criterion for a clear split of concordant and discordant genes.
After revisiting the PR method for the AGEMAP project, Owen found that
it is sensitive to consistent left- or right-sided departures. The PR
method is still easily dominated by one or two exceptionally
significant $p$-values, and does not identify which studies are
significant in distinguishing between concordant versus discordant
patterns (see first two examples in Table~\ref{5examples}).
\item[4.] Several forms of penalized or moderated $t$-statistics have been
proposed and shown to outperform traditional $t$-statistics [Efron et
al. (\citeyear{Efronh2001}); Tusher, Tibshirani and Chu (\citeyear{Tusherh2001}); Smyth (\citeyear{Smythh2004})]. For our algorithm we choose the
penalized t-statistics used in Efron et al. (\citeyear{Efronh2001}) and Tusher, Tibshirani and Chu (\citeyear{Tusherh2001}). The fudge
parameter $s_0$ is chosen to be the median variability estimator in the
genome.
\end{enumerate}
\end{rem*}

%s4 ###
\section{Applications}\label{sec:application}

%s4.1 ###
\subsection{Simulation study}\label{sec:simulation}
We conducted a simulation study for combining four data sets to compare
the performance among our proposed AW test, Fisher's EW test, Tippett's
minP method, Wilkinson's maxP method and Pearson's statistic (PR). For
each data set, we simulated five normal samples from a standard normal
distribution and five case samples from $ N(\theta,1)$. A total of
$g_1$ genes (category~I) were differentially expressed across all four
data sets; $g_2=400-g_1$ genes were differentially expressed in the
fourth data set only (category II); and 1600 genes were considered
null. Genes are called significant by controlling FDR at 5\% for each
method. Each simulation scenario was repeated 1000 times.

%t3 ###
\begin{table}[b]
% \centering
 %\begin{minipage}{160mm}
 \tablewidth=325pt
   \caption{Evaluation of AW, EW, minP, maxP and PR methods by simulations in the first scenario
   (I.~0 common DE genes; II. 400 4th-data set-specific DE genes; Null. 1600 random noise genes).
   Average number of genes detected in each category and the average FDR are shown under different effect size $\theta$}
\label{tab:0_400}
\begin{tabular}{@{}lcd{2.1}cccd{3.1}cc@{}}
  \hline
   & \multicolumn{4}{c}{$\bolds{\theta=2.0}$}&\multicolumn{4}{c@{}}{$\bolds{\theta=2.5}$}\\[-5pt]
& \multicolumn{4}{c}{\hrulefill}&\multicolumn{4}{c@{}}{\hrulefill}\\
  \textbf{Methods}& \textbf{I} & \multicolumn{1}{c}{\textbf{II}}  & \textbf{Null} &\textbf{ FDR (s.e.)}& \textbf{I} & \multicolumn{1}{c}{\textbf{II}}  & \textbf{Null} & \textbf{FDR (s.e.)}\\
  \hline
  AW  &0.0  &32.1 &1.9 &\phantom{0}4.8\% (0.002)
      &0.0  &137.1 &7.5 &\phantom{0}4.9\% (0.001)\\
  EW  &0.0  &7.6 &0.4 &\phantom{0}4.1\% (0.003)
      &0.0  &43.1 &2.0 &\phantom{0}4.2\% (0.002)\\
  minP  &0.0  &41.6 &2.4 &\phantom{0}5.0\% (0.002)
      &0.0  &163.0 &8.7 &\phantom{0}4.9\% (0.001)\\
  maxP  &0.0  &0.2 &0.1 &25.5\% (0.013)
      &0.0  &0.2 &0.1 &25.5\% (0.013)\\
  PR  &0.0  &3.2 &0.1 &\phantom{0}3.7\% (0.004)
      &0.0  &15.2 &0.4 &\phantom{0}2.2\% (0.002)\\
  \hline
\end{tabular}
\end{table}

%t4 ###
\begin{table}
% \centering
% \begin{minipage}{160mm}
\tablewidth=342pt
   \caption{Evaluation of AW, EW, minP, maxP and PR methods by simulations in the second scenario (I.~200 common DE genes; II. 200 4th-data set-specific DE genes;
   Null. 1600 random noise genes). Average number of genes detected in each category and the average FDR are shown under different effect size $\theta$}
\label{tab:200_200}
\begin{tabular}{@{}ld{3.1}d{2.1}d{2.1}ccd{3.1}d{2.1}c@{}}
  \hline
   & \multicolumn{4}{c}{$\bolds{\theta=1.5}$}&\multicolumn{4}{c@{}}{$\bolds{\theta=2.0}$}\\[-5pt]
& \multicolumn{4}{c}{\hrulefill}&\multicolumn{4}{c@{}}{\hrulefill}\\
  \textbf{Methods}& \multicolumn{1}{c}{\textbf{I}} & \multicolumn{1}{c}{\textbf{II}}  & \multicolumn{1}{c}{\textbf{Null}} & \textbf{FDR (s.e.)}& \textbf{I} & \multicolumn{1}{c}{\textbf{II}}  & \multicolumn{1}{c}{\textbf{Null}} & \textbf{FDR (s.e.)}\\
  \hline
  AW  &169.1  &24.3 &10.1 &4.9\% (0.0005)
      &198.7  &59.4 &13.4 &4.9\% (0.0004)\\
  EW  &188.4  &16.9 &8.5 &4.0\% (0.0004)
      &199.8  &35.4 &9.5 &3.9\% (0.0004)\\
  minP  &25.4  &6.9 &1.9 &5.0\% (0.0016)
      &144.0  &54.7 &10.3 &4.9\% (0.0005)\\
  maxP  &168.3  &3.7 &8.4 &4.6\% (0.0005)
      &195.7  &4.4 &9.8 &4.7\% (0.0005)\\
  PR  &178.7  &9.4 &3.8 &2.0\% (0.0003)
      &199.3  &21.3 &4.3 &1.9\% (0.0003)\\
  \hline
\end{tabular}
\end{table}

%t5 ###
\begin{table}[b]
% \centering
% \begin{minipage}{160mm}
   \caption{Evaluation of AW, EW, minP, maxP and PR methods by simulations in the third scenario (I. 400 common DE genes; II. 0 4th-data set-specific DE genes;
   Null. 1600 random noise genes). Average number of genes detected in each category and the average FDR are shown under different effect size $\theta$}
\label{tab:400_0}
\begin{tabular}{@{}ld{3.1}d{1.1}d{2.1}ccd{3.1}d{2.1}c@{}}
  \hline
   & \multicolumn{4}{c}{$\bolds{\theta=1.5}$}&\multicolumn{4}{c@{}}{$\bolds{\theta=2.0}$}\\[-5pt]
& \multicolumn{4}{c}{\hrulefill}&\multicolumn{4}{c@{}}{\hrulefill}\\
  \textbf{Methods}& \multicolumn{1}{c}{\textbf{I}} & \multicolumn{1}{c}{\textbf{II}}  & \multicolumn{1}{c}{\textbf{Null}} & \textbf{FDR (s.e.)}& \textbf{I} & \multicolumn{1}{c}{\textbf{II}}  & \multicolumn{1}{c}{\textbf{Null}} & \textbf{FDR (s.e.)}\\
  \hline
  AW  &359.3  &0.0 &18.6 &4.9\% (0.0004)
      &398.5  &0.0 &20.4 &4.8\% (0.0004) \\
  EW  &386.8  &0.0 &15.9 &4.0\% (0.0003)
      &399.8  &0.0 &16.1 &3.9\% (0.0003)\\
  minP  &121.3  &0.0 &6.3 &4.8\% (0.0007)
      &329.5  &0.0 &16.8 &4.8\% (0.0004) \\
  maxP  &357.5  &0.0 &19.0 &5.0\% (0.0004)
      &394.9  &0.0 &21.3 &5.1\% (0.0004)\\
  PR  &373.9  &0.0 &7.5 &2.0\% (0.0002)
      &399.4  &0.0 &7.8 &1.9\% (0.0002) \\
  \hline
\end{tabular}
\end{table}

%t6 ###
\begin{sidewaystable}
\tablewidth=551pt
\caption{Five genes from the mouse energy metabolism data. Moderated
$t$-statistics and $p$-values for individual studies are listed. $w*$
represents AW-obtained weight. AW2 represents AW concordant method}
\label{5examples}
\begin{tabular*}{\textwidth}{@{\extracolsep{\fill}}lccccccccc@{}}
\hline & \multicolumn{3}{c}{\textbf{Moderated $\bolds{t}$-statistic ($\bolds{p}$-value)}}& \multicolumn{5}{c}{\textbf{Is it detected $\bolds{(q(V)\leq5\%)}$?}}&\\[-5pt]
& \multicolumn{3}{c}{\hrulefill}& \multicolumn{5}{c}{\hrulefill}&\\
\textbf{Gene} &\textbf{Brown fat} &\textbf{Liver} &\textbf{Heart}  &\textbf{EW}  & \textbf{minP} & \textbf{PR}& \textbf{AW} & \textbf{AW2}& \textbf{Concordant?}\\
    %&&&&&& &&&(concordance)\\
\hline
  1423407\_a\_at   &2.2 & 1.7 & $-$3.7 & \\
      & (0.0027) & (0.0027) & (0.0014)  &$\surd$& $\times$ & $\surd$& $\surd$& $\times$ &no  \\
  $w*$         &1&1&1\\[3pt]
 1418429\_at   &3.6 & 1.1 & $-$3.2 & \\
      & (0.0003) & (0.067) & (0.002)  &$\surd$& $\times$ & $\surd$& $\surd$&$\times$&no  \\
  $w*$         &1&0&1\\[3pt]
1449015\_at   &0.4 & $-$3.3 & $-$1.8 & \\
      & (0.46) & (0.0009) & (0.011)  &$\surd$& $\times$ & $\surd$& $\surd$&$\surd$ &yes \\
  $w*$         &0&1&1\\[3pt]
1416415\_a\_at   &$-$0.8 & 2.2 & 2.6 & \\
      & (0.15) & (0.0026) & (0.0023)  &$\surd$& $\times$& $\surd$& $\surd$& $\surd$ &yes \\
  $w*$         &0&1&1\\[3pt]
1415727\_at   &$-$1.5 & $-$1.6 & $-$3.5 & \\
      & (0.018) & (0.014) & (0.0008)  &$\surd$& $\times$& $\surd$& $\surd$& $\surd$ &yes  \\
  $w*$         &1&1&1\\
  \hline
\end{tabular*}
\end{sidewaystable}

Summaries of the resulting FDR and average number of genes identified
in each category under three different scenarios appear in the
following tables: 0 category~I and 400 category II genes in Table
\ref{tab:200_200}; 200 category I and 200 category II genes in Table
\ref{tab:400_0}; 400 category I and 0 category II genes in Table
\ref{tab:0_400}. The results are consistent with the power calculation
discussed in Section~\ref{sec:power}. In Table~\ref{tab:0_400}, minP is
much more powerful than EW. When $\theta=2$, minP correctly detects an
average of 41.6 genes and EW detects only 7.6 genes.
AW detects 32.1
genes, considerably close to minP. Similarly, in Table~\ref{tab:400_0},
EW (386.8 genes are detected when $\theta=1.5$) is more powerful than
minP (121.3 genes detected) and AW (359.3 genes detected) is close to
EW in performance. Overall, AW performance was stable in these extreme
situations. We note most methods show FDR close to 5\%, although maxP
loses so much power at scenario 1 that FDR is inflated and the PR
method appears slightly conservative.

%s4.2 ###
\subsection{Energy metabolism in mouse model}\label{sec:mouse_energy}

An energy metabolism disorder in children is associated with very
longchain acyl-coenzyme A dehydrogenase (\mbox{VLCAD}) deficiencies. In an
ongoing unpublished project, two genotypes of the mouse model---wild
type (VLCAD $+$/$+$) and VLCAD-deficient (\mbox{VLCAD $-$/$-$})---were studied for
three types of tissues (brown fat, liver and heart) with 4 mice in each
genotype group. Microarray experiments were applied separately to study
the expression changes across genotypes. In this study we tested the
hypotheses that tissue-specific physiology triggers tissue-dependent
responses, with precise pools of differentially expressed genes
specific to the tissue in question. The purpose of this hypothesis is
to identify signature genes that are significant for tissue subsets---an analysis that corresponds to $\mathit{HS}_B$.

%f1 ###
\begin{figure}

\includegraphics{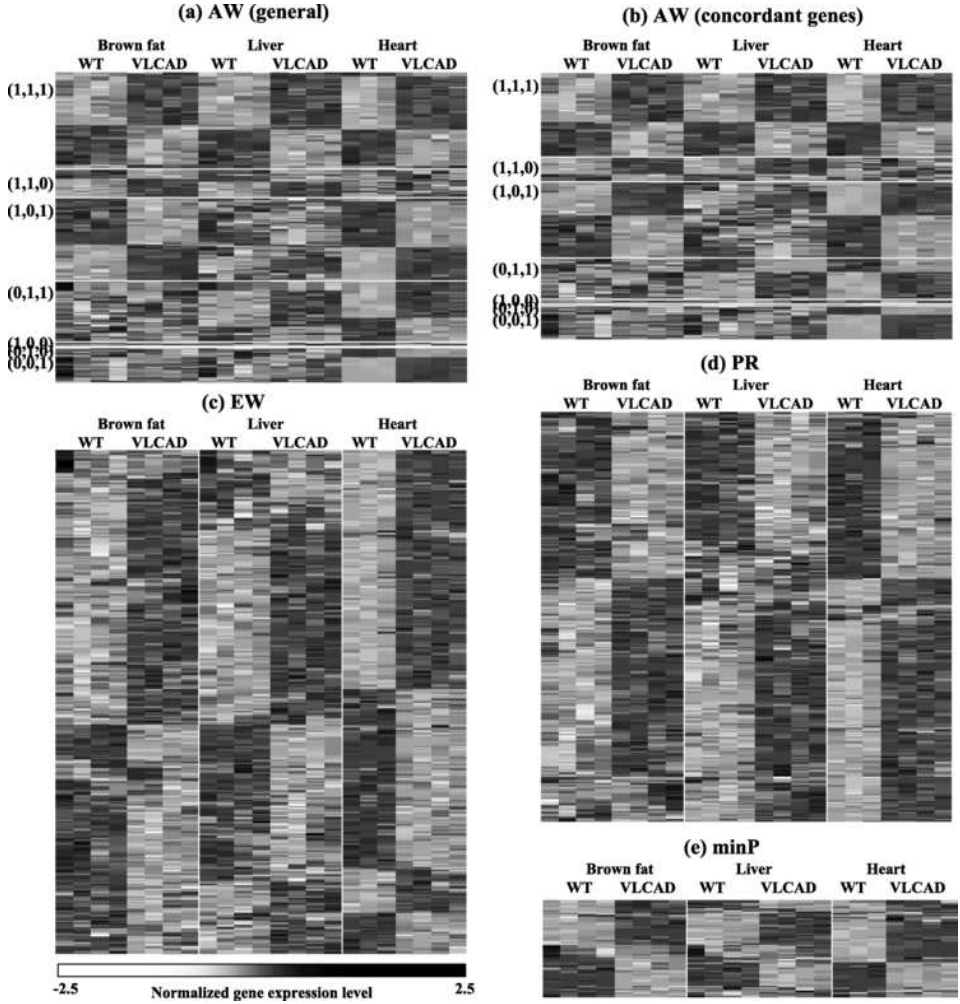}

\caption{Heatmaps of gene expressions for differentially expressed
genes identified by different methods in the mouse energy metabolism
data sets.} \label{fig:mouse}
\end{figure}

Due to the low power of maxP, the Figure~\ref{fig:mouse} data are
limited to AW, EW, minP and PR methods. Note that EW, minP and AW are
based on the summarization of $p$-values across studies, and that the
methods alone do not distinguish among discordant genes with difference
in expression across studies (e.g., up-regulated in one study but
down-regulated in another). The modified algorithm of AW for filtering
out discordant genes (Section~\ref{sec:AW}, step IV) can be implemented
in such situations, since it discards all discordant genes among
studies that contribute to the adaptive weight. The modified AW
algorithm is not applicable to EW, minP and PR because those methods do
not provide which studies should be considered for
concordance/discordance evaluations.

Overall, the general AW detects 203 genes [Figure~\ref{fig:mouse}(a)];
among these, 28 genes were conflicting in terms of up- or down-regulations---for example, Figure~\ref{fig:mouse}(b) shows the
detection of 175 genes. Adaptive-weights serve as a natural grouping
process for identified genes: 55 genes with weights of ($1,1,1$) are
differentially expressed in all three tissue types [Figure
\ref{fig:mouse}(b)], and 27 with weights of ($0,1,1$) were differentially
expressed in liver and heart tissues, but not in brown fat. The number
of detected genes related to heart tissue [($1,1,1$), ($1,0,1$), ($0,1,1$)
and ($0,0,1$) in Figure~\ref{fig:mouse}(b)] is much higher than that
related to brown fat or liver tissues, representing increase impact of
VLCAD deletion in heart metabolism activities. According to the EW
results shown in Figure~\ref{fig:mouse}(c), that method detected more
genes (329) than our proposed AW method. However, the identified gene
list is difficult to interpret and investigate, even after reordering
by hierarchical clustering. In this application minP appears to be much
less powerful.

To illustrate AW performance in terms of genes that consistently
regulate in the same direction across data sets, details for five genes
are presented in Table~\ref{5examples}. Four of the five methods
identified the five example genes as differentially expressed (the
exception was minP).
The first two genes (1423407\_a\_at and
1418429\_at) clearly indicate discordant regulation with opposite
moderated $t$-statistics between brown fat and heart. Even though
Pearson's method (PR) was specifically designed to detect concordant
genes, it failed to achieve this goal in this particular situation. In
contrast, our proposed AW method uses a post-hoc approach (Section
\ref{sec:AW}, step IV) to filter out discordant genes. Such a post-hoc
procedure is not feasible for EW, minP or PR without indicating which
studies are differentially expressed. For example, in 1449015\_at and
1416415\_a\_at, the AW method with concordance filtering will still
identify them as concordant DE genes, even though regulation of the
nonsignificant study (brown fat) contradicts the two significant
studies.
The difference between AW and the natural tendency of
biologists to pick studies based on $p$-values obtained from individual
analysis is illustrated by the fifth gene, 1415727\_at, which produces
moderate signals for brown fat and liver tissue and a very strong
signal for heart tissue, to the degree that it can easily be ignored
for brown fat and liver following adjustment for multiple comparisons.
It is, in general, difficult to decide whether it is a ($0,0,1$)- or
($1,1,1$)-type of gene. The fact that this gene is moderately significant
in two studies and very significant in a third study enabled AW to
determine that combining results across all three studies gives the
best statistical significance and it should be a ($1,1,1$)-type of gene.

%s4.3 ###
\subsection{Prostate cancer and lung cancer studies}\label{sec:prostate}

We applied the AW, EW, minP and PR methods to three sets of prostate
cancer data and three sets of lung cancer data (Table
\ref{three-datasets}). Some of the studies were performed by cDNA
technology [Dhanasekaran et al. (\citeyear{Dhanasekaranh2001}), Luo et al. (\citeyear{Luoh2001}) and Garber et al. (\citeyear{Garberh2001})] while
others used Affymetrix oligo-based technology [Welsh et al. (\citeyear{Welshh2001}),
Bhattacharjee et al. (\citeyear{Bhath2001}) and Beer et al. (\citeyear{Beerh2002})]. Data set probes were matched
according to their Entrez IDs; the intensities of multiple probes
matching the same ID were averaged. For the prostate cancer data set,
comparisons were made between clinically localized cancer and benign
tissues. For the lung cancer data set we compared tissues from
adenocarcinoma patients with those from healthy donors.

%f2 ###
\begin{figure}

\includegraphics{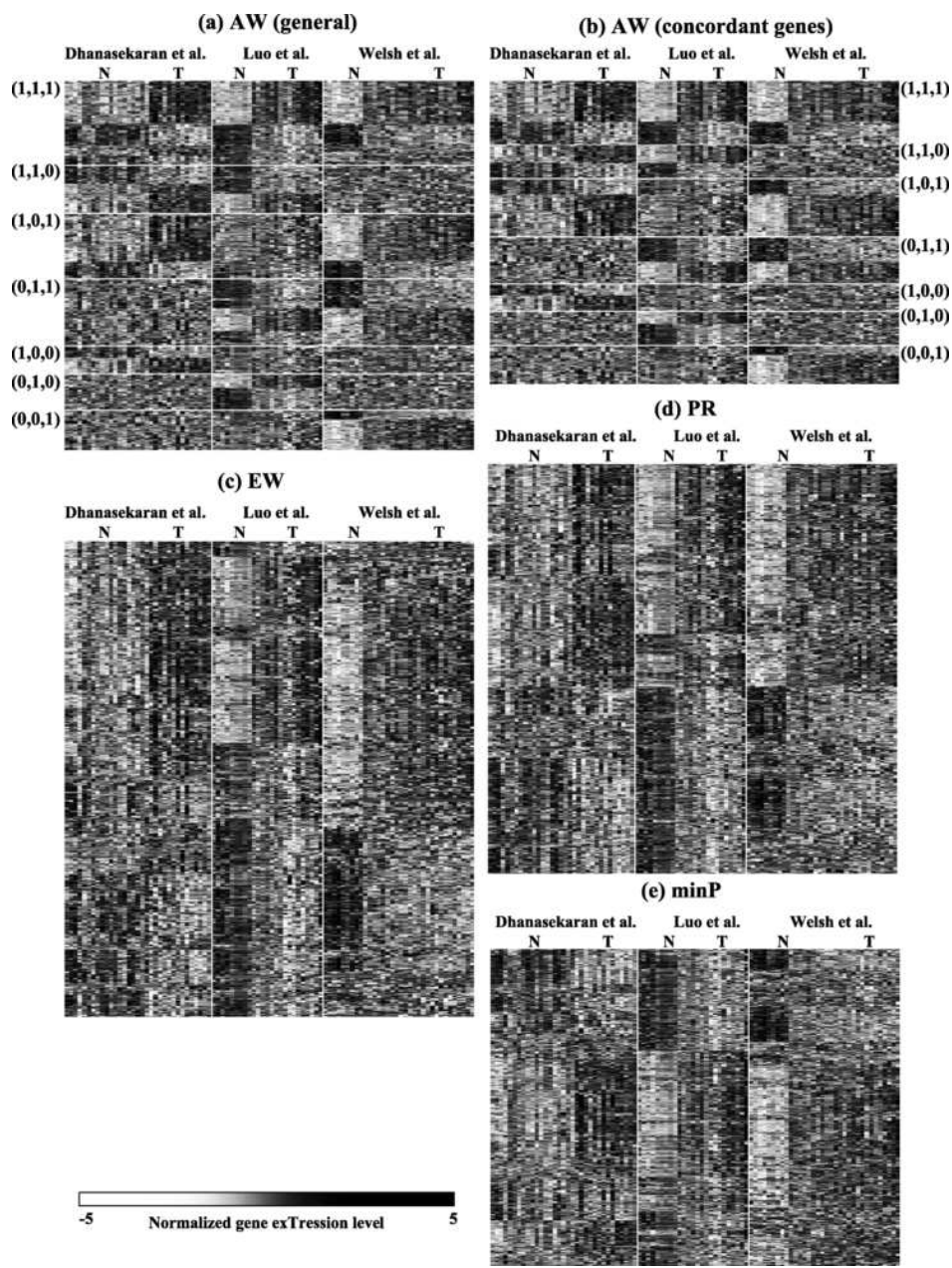}

\caption{Heatmaps of gene expression intensities for differentially
expressed genes identified by different methods in the prostate cancer
data sets.} \label{fig:prostate}
\end{figure}

\begin{figure}

\includegraphics{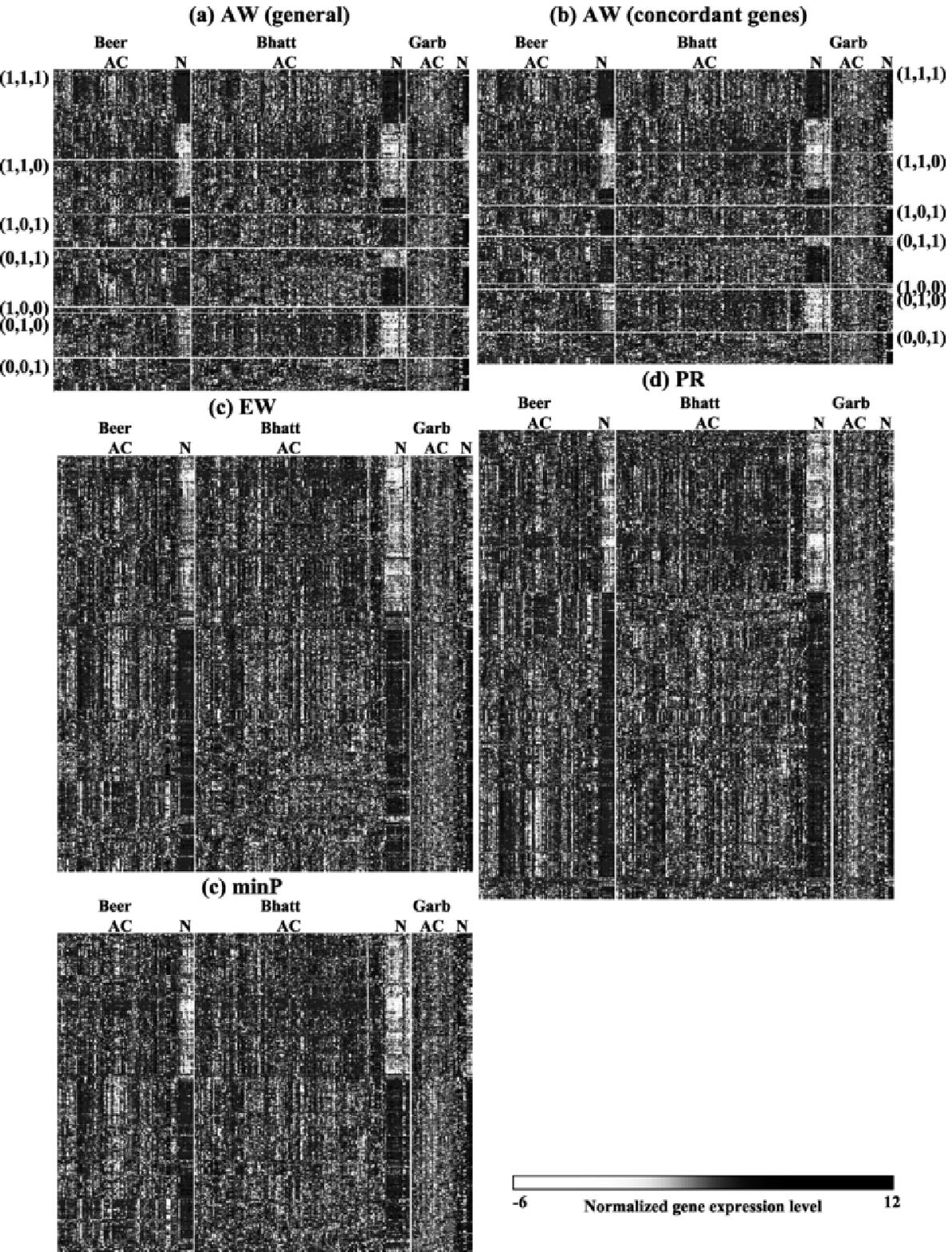}

\caption{Heatmaps of gene expression intensities for differentially
expressed genes identified by different methods in the lung cancer data
sets.} \label{fig:lung}
\end{figure}

The results shown in Figures~\ref{fig:prostate} and
\ref{fig:lung} reflect characteristics that are similar to those
discussed in the above mouse example. With an exception, minP did not
perform as poorly as it did in Section~\ref{sec:simulation}. Compared
to the other methods, our proposed AW method identified much clearer
patterns. Of the 722 genes in Figure~\ref{fig:prostate}(a), 618 genes
show consistent regulation across studies [Figure~\ref{fig:prostate}(b)].
Approximately 14\% of the identified genes were discordant across
studies. Possible causes of discordant genes may include mistaken gene
annotations in old array platforms [Dai et al. (\citeyear{Daih2005})],
differential probe efficiencies, heterogeneous sample populations
across studies and nonspecific cross hybridizations. According to our
findings, only moderately concordant information existed across the
three prostate cancer studies, probably because (a) their sample sizes
were small, or (b)~they entailed in-house cDNA arrays or commercial
products that were still in the early stages of development. Of the 618
concordant AW-detected genes, 130 genes (21\%) were consistent
($1,1,1$)-type biomarkers and 205 genes (33.2\%) were specific to one
study only: 55 ($1,0,0$)-type biomarkers, 70 of the ($0,1,0$) type, and 80
of the ($0,0,1$) type. The EW, minP and PR methods all detected slightly
greater numbers of biomarkers than the AW method (924, 745 and 882,
resp.). However, in each case the detected biomarkers were
difficult to interpret and follow up, and all three methods presented
challenges in terms of guaranteeing the detection of concordant genes
only. In summary, our findings suggest that results from individual
microarray studies require careful interpretation, and that integrative
analyses are appropriate as a validation tool.\looseness=-1

%f3 ###

Similar patterns and results were obtained when the four methods were
applied to lung cancer studies (Figure~\ref{fig:lung}). The AW method
detected 366 genes, with 349 confirmed as concordant (only 4.6\% are
discordant compared to 14.4\% in prostate cancer). Among the 349
concordant biomarkers, 99 were type ($1,1,1$) (28.4\% compared to 21\% in
prostate cancer) and 96 were single study specific (27.5\% compared to
33.2\% in prostate cancer): 7 type ($1,0,0$), 51 type ($0,1,0$) and 38 type
($0,0,1$). Overall, our lung cancer studies had more biomarkers that were
consistent in terms of concordant up-regulation and down-regulation
patterns, and fewer single study-specific biomarkers. These results
match those from previous reports showing better consistency among lung
cancer studies compared to prostate cancer studies, possibly due to
larger sample sizes, better gene annotations, more specific disease
subtype comparisons and better array quality. For example,
Bhattacharjee and Beer used Affymetrix platforms, while Garber's data
were generated from the lab of Pat Brown, the inventor of cDNA arrays.

%s5 ###
\section{Power and admissibility}\label{sec:compare}
In this section we drop the subscript $g$ for genes and assume
independence among studies when comparing five test statistics (EW, AW,
minP, maxP and PR) for $H_B$ at the univariate level. The maxP
statistic is included for demonstration purposes although it is not
targeted to $H_B$. To date, no best method for combining multiples
studies has been identified, therefore, choosing a combined statistic
must reflect specific biological purposes. Birnbaum (\citeyear{Birnbaumh1954,Birnbaumh1955})
established general conditions for evaluating combined methods,
including monotonicity and admissibility. To compare several combined
test procedures, he considered a one-sample test of the mean of a
Gaussian distribution with known variance. We will use a similar
two-sample test of the means of two Gaussian distributions with known
variance:
%e5.1 ###
\begin{equation}\label{eqn_Zk}
Z_k=\frac{\overline{X}_{2k}-\overline{X}_{1k}}{\sigma_k\sqrt{1/n_{k1}+1/n_{k2}}},\qquad
k=1,2,\ldots,K,
\end{equation}
where $\overline{X}_{1k}=(1/n_{k1})\cdot \sum_{s=1}^{n_{k1}} X_{ks}$,
$\overline{X}_{2k}=(1/n_{k2})\cdot \sum_{s=n_{k1}+1}^{n_{k1}+n_{k2}}
X_{ks}$, $X_{ks}\sim N(0, \break\sigma_k^2)$ when $1\leq s\leq n_{k1}$ and
$X_{ks}\sim N(\theta_k, \sigma_k^2)$ when $n_{k1}+1\leq s\leq
n_{k1}+n_{k2}$. We will use the two-sided $p$-values $P_k=\operatorname{Pr}(|Z|\geq
|z_k||\theta_k=0)$ for study $k$, where $Z$ is the standard normal
distribution, to examine the acceptance regions of the various combined
test procedures. The simplified framework is the focus for the
discussion in the \hyperref[sec:admissibility]{Appendix} of admissibility and power comparisons of
the five statistics. It is shown there that AW, EW, PR and minP are all
admissible, but maxP is not.

%s5.1 ###
\subsection{Power comparison of EW, AW, minP, maxP and PR under $H_{Bh'}$}\label{sec:power}

Denote by $\Theta_0=\{\theta_1=\cdots=\theta_K=0\}$ and
$\Theta_A=\{\mbox{at least one } \theta_k \neq 0\}$ (i.e., $H_B$) the
null and alternative hypothesis. Letting $\beta^{\mathrm{AW}}(\theta;\alpha)$ be
the power of a test controlled at level $\alpha$ for the OW statistic
given $\theta\in\Theta_A$, we have
%e5.2 ###
      \begin{equation}
      \qquad\beta^{\mathrm{AW}}(\theta;\alpha)=\operatorname{Pr}(V^{\mathrm{AW}}\leq C_{\alpha}^{\mathrm{AW}}|\theta)=
      1-\int_{\Omega^{\mathrm{AW}}}\prod_{k=1}^Kp(P_k|\theta)\,dP_1\,\cdots \,dP_K,
      \end{equation}
where $C_{\alpha}^{\mathrm{AW}}$ is the solution of $v$ to the equation
$P(V^{\mathrm{AW}}\leq v|\Theta_0)=\alpha$, $\Omega^{\mathrm{AW}}=
\bigcap_{j=1}^{2^K-1}\{p(u(w_j))>C_{\alpha}^{\mathrm{AW}}\}=\bigcap_{j=1}^{2^K-1}\{U(w_j)<F^{-1}_{\operatorname{Gamma}(\sum_{k=1}^K
w_{jk},1)}(1-C_{\alpha}^{\mathrm{AW}})\}$ and $F^{-1}_{\operatorname{Gamma}(\alpha,\beta)}$ is
the inverse CDF of a Gamma distribution with parameters $\alpha$ and
$\beta$, $w_j=(w_{j1},\ldots,w_{jK})$, $w_{jk}\in\{0,1\},k=1,\ldots,K$,
and enumeration index $j$ exhausts all different weight vector
possibilities such that $\sum_{k=1}^K w_{jk}\geq 1$. If the null
hypothesis is true, it is generally accepted that the individual $P_k$
value is uniformly distributed on [$0,1$]. The density of the $p$-value
under alternative law is expressed as
%e5.3 ###
  \begin{equation}
  p(P|\theta)={\frac{p(x|\theta)}{p(x|0)}}\biggl|_{ x=g(P)}\qquad (0\leq P \leq 1),
  \end{equation}
  where $x=g(P)$ indicates the solution of $P=\int_{x}^1f(x|0)\,dx$ [Pearson (\citeyear{Pearsonh1938})].
Similarly, the power for EW and minP can be calculated by
$\beta^{\mathrm{EW}}(\theta;\alpha)=
\int_{\Omega^{\mathrm{EW}}}\prod_{k=1}^Kp(P_k|\theta)\,dP_1\,\cdots\, dP_K$,
$\beta^{\mathrm{minP}}(\theta;\alpha)=1-[\int^1 _{C^{\mathrm{minP}}_{\alpha}}
p(P \mid \theta)\,dP]^K$ \vspace*{-2pt}and $\beta^{\mathrm{maxP}}( \theta  ; \alpha )=[\int^{C^{\mathrm{maxP}}_{\alpha}}_{\alpha}
p(P \mid \theta)\,dP]^K$, where $\Omega^{\mathrm{EW}}=\{-\sum_{k=1}^K \log p_k\geq C_{\alpha}^{\mathrm{EW}}\}$, $C_{\alpha}^{\mathrm{EW}}
=F_{\operatorname{Gamma}(K, 1)}^{-1}(1-\alpha)$, $C_{\alpha}^{\mathrm{minP}}=F_{\operatorname{Beta}(1,K)}^{-1} (\alpha)=1-(1-\alpha)^{1/K}$,
$C^{\mathrm{maxP}}_{\alpha} =\alpha^{1/K}$.

In our simplified setting, the $Z$ test in (3) is used for power
calculations, hence, the density of $P_k$ is
\begin{eqnarray}
  p(P_k|\theta_k)
  &=&
   \frac{1}{2}\exp\biggl\{\frac{c_k}{2}[2\Phi^{-1}(1-P_k/2)-c_k]\biggr\}\nonumber
   \\[-8pt]\\[-8pt]
   &&{}+\frac{1}{2}\exp\biggl\{-\frac{c_k}{2}[2\Phi^{-1}(1-P_k/2)+c_k]\biggr\},\nonumber
\end{eqnarray}
where $c_k=\frac{\theta_k}{\sigma_k\sqrt{1/n_{k1}+1/n_{k2}}}$,
$k=1,\ldots,K$. We consider\vspace*{1pt} $n_{k1}=n_{k2}=5$ and $\sigma_k=1$ so that
the effect size is represented by $\theta_k$ and power is evaluated
with varying effect sizes.

\begin{figure}

\includegraphics{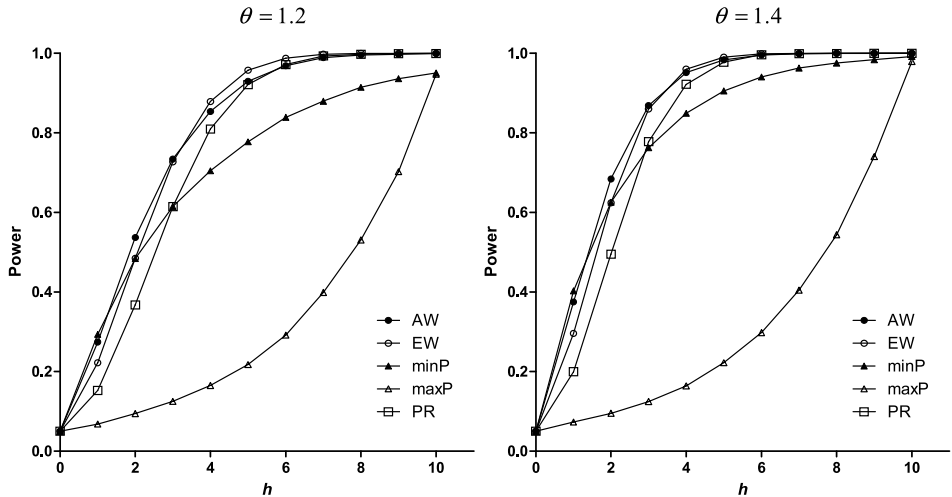}

\caption{Power analysis of EW, AW, minP, PR and maxP under $H_{Bh'}$,
$1\leq h\leq K$. We compare power curves of the five methods combining
$K=10$ studies. X axis represents $h$, the number of studies that have
nonzero effects. }\label{fig_power}
\end{figure}

The graphs in Figure~\ref{fig_power} reflect a situation in which
$K=10$ for simplified alternative hypothesis $H_{Bh'}$ ($1\leq h\leq
K$). Studies with nonzero effect sizes share a common effect size
$\theta$. Power curves under $\theta\in \{1.2, 1.4\}$ and varying
values of~$h$ are displayed. Due to the difficulty of achieving an
exact power calculation for $K=10$, we performed 10,000 simulations to
generate power curves. EW and AW are calculated for one-sided $p$-values
for the purpose of comparability with PR, maxP, minP. In application,
it is unlikely that the signs of effect will be known, therefore,
two-sided $p$-values for maxP, minP, EW and AW are preferred. As
expected, the figure shows that minP is more powerful than EW when $h$
is small, and EW is more powerful than minP when $h$ is large. On the
other hand, AW performs stably and comparably to the best method in
situations involving the two extremes. The performance of maxP further
confirms Loughin's conclusion that it has very low power unless $h=K$.

%f4 ###

%s6 ###
\section{Discussion}\label{sec:discussion}
In this paper we described our proposal for an adaptively weighted (AW)
statistic for combining multiple studies, and reported our findings
after applying it to two sets of combined microarray studies.
Acknowledging that meta-analysis methods depend heavily on the
biological question being investigated, we formulated two statistical
hypothesis settings ($\mathit{HS}_A$ and $\mathit{HS}_B$) to identify differentially
expressed genes considered significant in either partial or full data
sets. Classical EW, minP and our proposed AW methods were used to
analyze $\mathit{HS}_A$.

\begin{figure}

\includegraphics{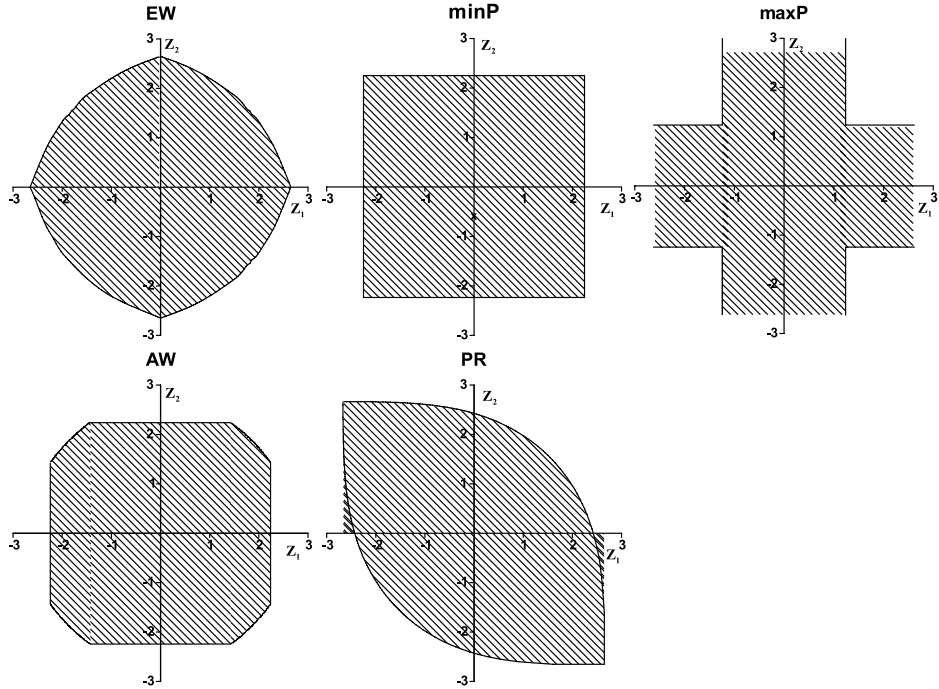}

\caption{Acceptance regions of EW, AW, minP, PR and maxP statistic for
combining $p$-values from two independent studies when testing means of
Gaussian distributions with known variances. }\label{fig_acceptRegion}
\end{figure}

According to our findings, AW, EW and minP are all admissible in
simplified scenarios. In terms of power analysis, EW was more powerful
when all data sets were significant, while minP was more powerful when
only one or a small number of data sets were significant. As a
compromise between EW and minP, the AW method performed close to the
best method in either extreme alternative hypothesis setting (Figure
\ref{fig_acceptRegion}). Simulation results also confirmed this robust
property of AW (Tables~\ref{tab:0_400}--\ref{tab:400_0}). In
applications, AW had the additional advantage of categorizing
differentially expressed genes by their adaptive weights, thus
providing a practical basis for further biological exploration. In
addition to not detecting discordantly regulated genes, the modified
algorithm in Section~\ref{sec:AW}, step IV, was appealing for the
specific biological purpose of identifying all nondiscordant genes.

In this project we restricted the binary $0,1$ adaptive weight search
space for purposes of computational convenience and biological
interpretability. For example, in Figure~\ref{fig:mouse}(b) the AW data
support an immediate categorization of detected biomarkers, as well as
information on similar/dissimilar differential gene expression between
tissue pairs. As shown for the EW data in Figure~\ref{fig:mouse}(c),
Fisher's method generated a large number of nontraceable biomarkers
that were difficult to work with in terms of follow-up analyses.
Theoretically, it is possible to extend the $0,1$ space to a
nonrestricted real number (i.e., positive weights that add up to 1).
However, such results generate biomarker lists similar to those
generated by the EW method [Figure~\ref{fig:mouse}(c)]. In other words,
using nonbinary weights may be slightly superior statistically, but not
biologically.

%f5 ###

There are three limitations in addition to possible future extensions
for future research. First, we assumed that all studies contain an
identical matched gene list with no missing values. In actual practice,
separate studies to be combined usually come from different microarray
platforms. Requiring an identical matched gene list and no missing
values will exclude many important genes that appear in certain studies
but not in others, thus requiring an extension that allows for missing
values. Second, we focused on two-group comparison in this paper, and
made a modification in order to limit detection to genes with
concordant expression changes. To compare more than two groups, the
$F$-statistic and its variations can be applied; resulting $p$-values from
$F$ tests can be combined similarly as described for the algorithm in
Section~\ref{sec:AW}. However, small $p$-values across studies do not
guarantee concordant expression patterns. To address this problem, we
have developed a multi-class correlation approach [Lu, Li and Tseng
(\citeyear{Luh2010})]. Third, our proposed method focuses on $\mathit{HS}_B$ rather than
$\mathit{HS}_A$, which is not the case with many biological applications.
Finally, the AW statistic can be extended from biomarker detection to
gene set enrichment analyses. Note that post-meta-analysis enriched
pathways (gene sets) are thought to be more supportive of biological
interpretations.

While we only considered combining multiple microarray studies in this
paper, the methods we described can easily be extended to combinations
of multiple genomic, epigenomic and/or proteomic studies---for
instance, data sets from SNP arrays, genome arrays, methylation arrays,
proteomic experiments and ChIP-on-chip experiments. Additional
extensions and/or alternative models are required to accommodate
biological knowledge and to address specific questions of interest.

\begin{appendix}\label{sec:admissibility}

\section*{Appendix: Admissibility}

A test is considered admissible if it cannot be uniformly improved by
any other test. No single test has been accepted as the most powerful,
even in the simplified scenarios. Birnbaum expressed a necessary and
sufficient condition (known as Theorem 5.1) for any test to be
admissible under this situation.

\begin{theo}[\textup{[Birnbaum (\citeyear{Birnbaumh1954,Birnbaumh1955})]}]
Under $H_B$ and the test statistic is in the exponential family [e.g.,
equation (\ref{eqn_Zk})], the necessary and sufficient condition for a combined
test procedure to be admissible is that the corresponding acceptance
region is convex.
\end{theo}

Since the acceptance regions of EW and minP have been identified as
convex, both methods are admissible; maxP is not. When proving that the
PR method is admissible, Owen (\citeyear{Owenh2009}) clarified Birnbaum's (\citeyear{Birnbaumh1954})
misinterpretation of the PR method. The acceptance regions of EW, minP,
maxP, AW and PR on the plane of a pair of $Z$ statistics at level 0.05
are shown in Figure~\ref{fig_acceptRegion}. When illustrating the
rejection regions of several common combined tests (including EW and
minP), Birnbaum showed a preference because it appeared to be ``fairly
sensitive in all directions.'' From Figure~\ref{fig_acceptRegion}, it
is clear that the PR method prefers effects that show common directions
in two studies, since the rejection regions in the first and third
quadrants are less stringent than the second and fourth quadrants. Note
that AW actually shares positive aspects of both EW and minP methods:
generally more~sensitive than minP when parameters from both studies
depart from 0 and more sensitive than EW when only one of the
parameters departs from 0, and more sensitive than the minP method when
parameters from both studies depart from 0. According to the following
corollary, AW is admissible because the intersection of convex sets is
convex, therefore, its acceptance region is convex.

\begin{corollary}
The acceptance region of AW is convex and, thus, AW is admissible under
$H_B$ and assumption (\ref{eqn_Zk}).
\end{corollary}

\begin{pf}
Denote by $p_k=2(1-\Phi(|z_k|))$ the two-sided $p$-value, where
$\Phi(t)=\int_{-\infty}^t\phi(t)\,dt$, $\phi(t)$ is the density of the
standard normal distribution. First we prove that
$f(z_k)=-\log(p_k)=-\log(1-\Phi(|z_k|))+C$ is convex.
$f''(z)=\frac{\phi(|z|)}{[1-\Phi(|z|)]^2}\{\phi(|z|)-|z|[1-\Phi(|z|)]\}$
when $z\neq 0$. It is well known that the elementary upper bound for
$1-\Phi(x)$ is $\phi(x)/x$, for $x>0$. Thus, $f''(z)>0$ when $z\neq 0$.
Since $f(z)$ is continuous at $z=0$, $f(z)$ is convex in $z$. Hence,
$f(z_1,z_2,\ldots,z_n)=-\sum_{k=1}^n \log(p_k)$ for any $n\geq 1$ is
convex, because the sum of convex functions is convex. For the AW
statistic, the acceptance region is $\{z_1,z_2,\ldots,z_K:\min_{1\leq k
\leq K}p(u(w))>c\}$, where $p(u(w))$ is the right-sided $p$-value of
$U(w)$:
\begin{eqnarray*}
&&
\Bigl\{z_1,z_2,\ldots,z_K\dvtx\min_{0\leq k \leq K}p(u(w))>c\Bigr\}
\\
&&\qquad=
\bigcap_{I_k\in \{0,1\},1\leq k \leq K}\Biggl\{z_1,z_2,\ldots,z_K:p\Biggl(-\sum_{k=1}^K \log[p_k^{I_k}]\Biggr)>c\Biggr\}
\\
&&\qquad=
\bigcap_{I_k\in \{0,1\},1\leq k \leq K}\Biggl\{z_1,z_2,\ldots,z_K\dvtx -\sum_{k=1}^K \log[p_k^{I_k}]<\gamma_j\Biggr\},
\\
&&
\hspace*{30pt}\hphantom{\bigcap_{I_k\in \{0,1\},1\leq k \leq K}\Biggl\{z_1,z_2,\ldots,z_K\dvtx -\sum_{k=1}^K \log[p_k^{I_k}]<\gamma_j\Biggr\},}j=1,2,\ldots,2^K-1,
\end{eqnarray*}
$\gamma_j$ is $F^{-1}_{\operatorname{Gamma}(\sum_{k=1}^K I_k,1)}(1-c)$. Thus, the
acceptance region of AW is convex since the intersection of convex sets
is also convex.
\end{pf}
\end{appendix}

\section*{Acknowledgments}
The authors would like to thank Gerard
Vockley for providing the mouse metabolism data set and reviewers for
insightful comments.

\printaddresses


\begin{thebibliography}{99}

%b1 ###
\bibitem[\protect\citeauthoryear{}{2002}]{Beerh2002}
\textsc{Beer, D. G., Kardia, S. L., Huang, C. C., Giordano, T. J.,
Levin, A. M., Misek, D. E., Lin, L., Chen, G., Gharib, T. G., Thomas,
D. G., Lizyness, M. L., Kuick, R., Hayasaka, S., Taylor, J. M. G.,
Iannettoni, M. D., Orringer, M. B.} and \textsc{Hanash, S.} (2002).
Gene-expression profiles predict survival of patients with lung
adenocarcinoma. \textit{Nature Med.} \textbf{8} 816--824.

%b2 ###
\bibitem[\protect\citeauthoryear{}{1982}]{Bergerh1982}
\textsc{Berger, R. L.} (1982). Multiparameter hypothesis testing and
acceptance sampling. \textit{Technometrics} \textbf{24} 295--300.
\MR{0687187}

%b3 ###
\bibitem[\protect\citeauthoryear{}{2001}]{Bhath2001}
\textsc{Bhattacharjee, A., Richards, W. G., Staunton, J., Li, C.,
Monti, S., Vasa, P., Ladd,~C., Beheshti, J., Bueno, R., Gillette, M.,
Loda, M., Weber, G., Mark, E. J., Lander, E. S., Wong, W., Johnson, B.
E., Golub, T. R., Sugarbaker, D. J.} and \textsc{Meyerson, M.} (2001).
Classification of human lung carcinomas by mrna expression profiling
reveals distinct adenocarcinoma subclasses. \textit{Proc. Natl. Acad.
Sci. USA} \textbf{98} 13790--13795.

%b4 ###
\bibitem[\protect\citeauthoryear{}{1954}]{Birnbaumh1954}
\textsc{Birnbaum, A.} (1954). Combining independent tests of
significance. \textit{J. Amer. Statist. Assoc.} \textbf{49} 559--574.
\MR{0065101}

%b5 ###
\bibitem[\protect\citeauthoryear{}{1955}]{Birnbaumh1955}
\textsc{Birnbaum, A.} (1955). Characterizations of complete classes of
tests of some multiparametric hypotheses, with applications to
likelihood ratio tests. \textit{Ann. Math. Statist.} \textbf{26}
21--36.
\MR{0067438}

%b6 ###
\bibitem[\protect\citeauthoryear{}{2005}]{Boroveckih2005}
\textsc{Borovecki, F., Lovrecic, L., Zhou, J., Jeong, H., Then, F.,
Rosas, H. D., Hersch, S. M., Hogarth, P., Bouzou, B., Jensen, R. V.}
and \textsc{Krainc D.} (2005). Genome-wide expression profiling of
human blood reveals biomarkers for Huntington's disease. \textit{Proc.
Natl. Acad. Sci. USA} \textbf{102} 11023--11028.

%b7 ###
\bibitem[\protect\citeauthoryear{}{2007}]{Cardosoh2007}
\textsc{Cardoso, J., Boer, J. H., Morreau, H.} and \textsc{Fodde, R.}
(2007). Expression and genomic profiling of colorectal cancer.
\textit{Biochim. Biophys. Acta Rev. Cancer} \textbf{1775} 103--137.

%b8 ###
\bibitem[\protect\citeauthoryear{}{2007}]{Choih2007}
\textsc{Choi, H., Shen, R., Chinnaiyan, A. M.} and \textsc{Ghosh, D.}
(2007). A latent variable approach for meta-analysis of gene expression
data from multiple microarray experiments. \textit{BMC Bioinformatics}
\textbf{8} 364--383.

%b9 ###
\bibitem[\protect\citeauthoryear{}{2003}]{Choih2003}
\textsc{Choi, J. K., Yu, U., Kim, S.} and \textsc{Yoo, O. J.} (2003).
Combining multiple microarray studies and modeling interstudy
variation. \textit{Bioinformatics} \textbf{19} 84--90.

%b10 ###
\bibitem[\protect\citeauthoryear{}{2007}]{Conlonh2007}
\textsc{Conlon, E. M., Song, J. J.} and \textsc{Liu, A.} (2007).
Bayesian meta-analysis models for microarray data: A comparative study.
\textit{BMC Bioinformatics} \textbf{8} 80--100.

%b11 ###
\bibitem[\protect\citeauthoryear{}{2006}]{Conlonh2006}
\textsc{Conlon, E. M., Song, J. J.} and \textsc{Liu, J. S.} (2006).
Bayesian models for pooling microarray studies with multiple sources of
replications. \textit{BMC Bioinformatics} \textbf{7} 247--250.

%b12 ###
\bibitem[\protect\citeauthoryear{}{2007}]{Cousinsh2007}
\textsc{Cousins, R. D.} (2007). Annotated bibliography of some papers on
combining significances or $p$-values. Available at
\href{http://arxiv.org/abs/0705.2209v1}{arXiv:0705.2209v1}.

%b13 ###
\bibitem[\protect\citeauthoryear{}{2005}]{Daih2005}
\textsc{Dai, M., Wang, P., Boyd, A. D., Kostov, G., Athey, B., Jones,
E. G., Bunney, W. E., Myers, R. M., Speed, T. P., Akil, H., Watson, S.
J.} and \textsc{Meng, F.} (2005). Evolving gene/transcript definitions
significantly alter the interpretation of genechip data.
\textit{Nucleic Acids Res.} \textbf{33} e175. doi:
\href{http://dx.doi.org/10.1093/nar/gni179}{10.1093/nar/gni179}.

%b15 ###
\bibitem[\protect\citeauthoryear{}{2001}]{Dhanasekaranh2001}
\textsc{Dhanasekaran, S. M., Barrette, T. R., Ghosh, D., Shah, R.,
Varambally, S., Kurachi, K., Pienta, K. J., Rubin, M. A.} and
\textsc{Chinnaiyan, A. M.} (2001). Delineation of prognostic biomarkers
in prostate cancer. \textit{Nature} \textbf{412} 822--826.

%b17 ###
\bibitem[\protect\citeauthoryear{}{2001}]{Efronh2001}
\textsc{Efron, B., Tibshirani, J. D., Storey, R.} and
\textsc{Tusher, V.} (2001). Empirical Bayes analysis of a microarray
experiment. \textit{J. Amer. Statist. Assoc.} \textbf{96} 1151--1160.
\MR{1946571}

%b18 ###
\bibitem[\protect\citeauthoryear{}{1932}]{Fisherh1932}
\textsc{Fisher, R. A.} (1932). \textit{Statistical Methods for Research
Workers}, 4 ed. Oliver and Boyd, Edinburgh.

%b20 ###
\bibitem[\protect\citeauthoryear{}{2001}]{Garberh2001}
\textsc{Garber, M. E., Troyanskaya, O. G., Schluens, K., Petersen, S.,
Thaesler, Z., Pacyna-Gengelbach, M., van de Rijn, M., Rosen, G. D.,
Perou, C. M., Whyte,~R.~I., Altman, R. B., Brown, P. O., Botstein, D.}
and \textsc{Petersen, I.} (2001). Diversity of gene expression in
adenocarcinoma of the lung. \textit{Proc. Natl. Acad. Sci. USA}
\textbf{98} 13784--13789.

%b21 ###
\bibitem[\protect\citeauthoryear{}{1977}]{Georgeh1977}
\textsc{George, E. O.} (1977). Combining independent one-sided and
two-sided statistical tests---some theory and applications. Ph.D.
thesis, Univ. Rocheser.
\MR{2627130}

%b22 ###
\bibitem[\protect\citeauthoryear{}{2003}]{Ghoshh2003}
\textsc{Ghosh, D., Barrette, T. R., Rhodes, D.} and \textsc{Chinnaiyan,
A. M.} (2003). Statistical issues and methods for meta-analysis of
microarray data: A case study in prostate cancer. \textit{Functional
and Integrative Genomic} \textbf{3} 180--188.

%b23 ###
\bibitem[\protect\citeauthoryear{}{1955}]{Goodh1955}
\textsc{Good, I. J.} (1955). On the weighted combination of
significance tests. \textit{J. Roy. Statist. Soc. Ser. B} \textbf{17}
264--265.
\MR{0076252}

%b24 ###
\bibitem[\protect\citeauthoryear{}{1985}]{HedgesandOlkinh1985}
\textsc{Hedges, L. V.} and \textsc{Olkin, I.} (1985).
\textit{Statistical Methods for Meta-Analysis}. Academic Press, New York.
\MR{0798597}

%b26 ###
\bibitem[\protect\citeauthoryear{}{2006}]{Hongh2006}
\textsc{Hong, F., Breitling, R., McEntee, C. W., Wittner, B. S.,
Nemhauser, J. L.} and \textsc{Chory,~J.} (2006). Rankprod: A
bioconductor package for detecting differentially expressed genes in
meta-analysis. \textit{Bioinformatics} \textbf{22} 2825--2827.

%b27 ###
\bibitem[\protect\citeauthoryear{}{2005}]{Huh2005}
\textsc{Hu, P., Greenwood, C. M. T.} and \textsc{Beyene, J.} (2005).
Integrative analysis of multiple gene expression profiles with
quality-adjusted effect size models. \textit{BMC Bioinformatics}
\textbf{6} 128--138.

%b28 ###
\bibitem[\protect\citeauthoryear{}{1961}]{Lancasterh1961}
\textsc{Lancaster, H.} (1961). The combination of probabilities: An
application of orthonormal functions. \textit{Austr. J. Statist.}
\textbf{3} 20--33.
\MR{0130742}

%b29 ###
\bibitem[\protect\citeauthoryear{}{1971}]{Littleh1971}
\textsc{Littell, R. C.} and \textsc{Folks, J. L.} (1971). Asymptotic
optimality of Fisher's method of combining independent tests.
\textit{J. Amer. Statist. Assoc.} \textbf{66} 802--806.
\MR{0312634}

%b30 ###
\bibitem[\protect\citeauthoryear{}{1973}]{Littleh1973}
\textsc{Littell, R. C.} and \textsc{Folks, J. L.} (1973). Asymptotic
optimality of Fisher's method of combining independent tests, ii.
\textit{J. Amer. Statist. Assoc.} \textbf{68} 193--194.
\MR{0375577}

%b31 ###
\bibitem[\protect\citeauthoryear{}{2004}]{Loughinh2004}
\textsc{Loughin, T. M.} (2004). A systematic comparison of methods for
combining $p$-values from independent tests. \textit{Comput. Statist.
Data Anal.} \textbf{47} 467--485.
\MR{2086483}

%b32 ###
\bibitem[\protect\citeauthoryear{}{2010}]{Luh2010}
\textsc{Lu, S., Li, J., Song, C., Shen, K.} and \textsc{Tseng, G. C.}
(2010). Biomarker detection in the integration of multiple multi-class
genomic studies. \textit{Bioinformatics} \textbf{26} 333--340.

%b33 ###
\bibitem[\protect\citeauthoryear{}{2001}]{Luoh2001}
\textsc{Luo, J., Duggan, D. J., Chen, Y., Sauvageot, J., Ewing, C. M.,
Bittner, M. L., Trent, J. M.} and \textsc{Isaacs, W. B.} (2001). Human
prostate cancer and benign prostatic hyperplasia:  Molecular dissection
by gene expression profiling. \textit{Cancer Res.} \textbf{61}
4683--4688.

%b34 ###
\bibitem[\protect\citeauthoryear{}{2003}]{Moreauh2003}
\textsc{Moreau, Y., Aerts, S., De Moor, B., De Strooper, B.} and
\textsc{Dabrowski, M.} (2003). Comparison and meta-analysis of
microarray data: From the bench to the computer desk. \textit{Trends
Genet.} \textbf{19} 570--577.

%b35 ###
\bibitem[\protect\citeauthoryear{}{2001}]{OlkinandSanerh2001}
\textsc{Olkin, I.} and \textsc{Saner, H.} (2001). Approximations for
trimmed Fisher procedures in research synthesis. \textit{Statist.
Methods Med. Res.} \textbf{10} 267--276.

%b36 ###
\bibitem[\protect\citeauthoryear{}{2009}]{Owenh2009}
\textsc{Owen, A. B.} (2009). Karl Pearson's meta-analysis revisited.
\textit{Ann. Statist.} \textbf{37} 3867--3892.
\MR{2572446}

%b37 ###
\bibitem[\protect\citeauthoryear{}{2004}]{Parmigianih2004}
\textsc{Parmigiani, G., Garrett-Mayer, E. S., Anbazhagan, R.} and
\textsc{Gabrielson, E.} (2004). A~cross-study comparison of gene
expression studies for the molecular classificaiton of lung cancer.
\textit{Clin. Cancer Res.} \textbf{10} 2922--2927.

%b38 ###
\bibitem[\protect\citeauthoryear{}{1938}]{Pearsonh1938}
\textsc{Pearson, E. S.} (1938). The probability integral transformation
for testing goodness of fit and combining independent tests of
significance. \textit{Biometrika} \textbf{30} 134--148.

%b39 ###
\bibitem[\protect\citeauthoryear{}{1934}]{Pearsonh1934}
\textsc{Pearson, K.} (1934). On a new method of determining 'goodness
of fit.' \textit{Biometrika} \textbf{26} 425--442.

%b40 ###
\bibitem[\protect\citeauthoryear{}{2007}]{Piroozniah2007}
\textsc{Pirooznia, M., Nagarajan, V.} and \textsc{Deng, Y.} (2007).
Gene venn---a web application for comparing gene lists using venn
diagram. \textit{Binformation} \textbf{1} 420--422.

%b42 ###
\bibitem[\protect\citeauthoryear{}{2002}]{Rhodesh2002}
\textsc{Rhodes, D., Barrette, T. R., Rubin, M. A., Ghosh, D.} and
\textsc{Chinnaiyan, A. M.} (2002). Meta-analysis of microarrays:
Interstudy validation of gene expression profiles reveals pathway
dysregulation in prostate cancer. \textit{Cancer Res.} \textbf{62}
4427--4433.

%b43 ###
\bibitem[\protect\citeauthoryear{}{1953}]{Royh1953}
\textsc{Roy, S. N.} (1953). On a heuristic method of test construction
and its use in multivariate analysis. \textit{Ann. Math. Statist.}
\textbf{24} 220--238.
\MR{0057519}

%b44 ###
\bibitem[\protect\citeauthoryear{}{2004}]{Segalh2004}
\textsc{Segal, E., Friedman, N., Koller, D.} and \textsc{Regev, A.}
(2004). A module map showing conditional activity of expression modules
in cancer. \textit{Nature Genet.} \textbf{3} 1090--1098.

%b45 ###
\bibitem[\protect\citeauthoryear{}{2004}]{Shenh2004}
\textsc{Shen, R., Ghosh, D.} and \textsc{Chinnaiyan, A. M.} (2004).
Prognostic meta-signature of breast cancer developed by two-stage
  mixture modeling of microarray data.
\textit{BMC Genomics} \textbf{5} 94--109.

%b46 ###
\bibitem[\protect\citeauthoryear{}{2004}]{Smythh2004}
\textsc{Smyth, G. K.} (2004). Linear models and empirical Bayes methods
for assessing differential expression in microarray experiments.
\textit{Statist. Appl. Genet. Mol. Biol.} \textbf{3} Article 3.
\MR{2101454}

\bibitem[\protect\citeauthoryear{}{2002}]{s2002}
\textsc{Storey, J. D.} (2002). A direct approach to false discovery
rates. \textit{J. R. Stat. Soc. Ser. B} \textbf{64} 479--495.

%b47 ###
\bibitem[\protect\citeauthoryear{}{1949}]{Stoufferh1949}
\textsc{Stouffer, S., Suchman, E., DeVinnery, L., Star, S.} and
\textsc{Williams, J.} (1949). \textit{The American Soldier, Vol. I:
Adjustement during Army Life.} Princeton Univ. Press, Princeton, NJ.

%b48 ###
\bibitem[\protect\citeauthoryear{}{1931}]{Tippetth1931}
\textsc{Tippett, L. H. C.} (1931). \textit{The Methods in Statistics},
 1st ed. Williams and Norgate, London.

%b49 ###
\bibitem[\protect\citeauthoryear{}{2001}]{Tsengh2001}
\textsc{Tseng, G. C., Oh, M. K., Rohlin, L., Liao, J. C.} and
\textsc{Wong, W. H.} (2001). Issues in cdna microarray analysis:
Quality filtering, channel normalization, models of variations and
assessment of gene effects. \textit{Nucleic Acids Res.} \textbf{29}
2549--2557.

%b50 ###
\bibitem[\protect\citeauthoryear{}{2001}]{Tusherh2001}
\textsc{Tusher, V. G., Tibshirani, R.} and \textsc{Chu, G.} (2001).
Significance analysis of microarrays applied to the ionizing radiation
response. \textit{Proc. Natl. Acad. Sci. USA} \textbf{98} 5116--5121.\

%b51 ###
\bibitem[\protect\citeauthoryear{}{2001}]{Welshh2001}
\textsc{Welsh, J. B., Sapinoso, L. M., Su, A. I., Kern, S. G.,
Wang-Rodriguez, J., Moskaluk, C. A., Frierson, H. F.} and
\textsc{Hampton Jr., G. M.} (2001). Analysis of gene expression
identifies candidate markers and pharmacological targets in prostate
cancer. \textit{Cancer Res.} \textbf{61} 5974--5978.

%b52 ###
\bibitem[\protect\citeauthoryear{}{1951}]{Wilkinsonh1951}
\textsc{Wilkinson, B.} (1951). A statistical consideration in
psychological research. \textit{Psychol. Bull.} \textbf{48} 156--157.

\end{thebibliography}
\end{document}